\documentclass[aps,showpacs,manuscript,12pt]{revtex4}
\usepackage{amsfonts}
\usepackage{amssymb}
\usepackage{amsmath}
\usepackage{graphicx}

\begin{document}

\title{\textbf{On the behavior of homogeneous, isotropic and stationary
turbulence}}
\author{M. Tessarotto$^{1,2}$ and C. Asci$^{1}$}
\affiliation{$^{1}$Department of Mathematics and Informatics,\\
University of Trieste, Trieste, Italy \\
$^{2}$ Consortium for Magnetofluid Dynamics, Trieste, Italy }

\begin{abstract}
The recent development of a statistical model for incompressible
Navier-Stokes (NS) fluids based on inverse kinetic theory (IKT, 2004-2008)
poses the problem of searching for particular realizations of the theory
which may be relevant for \textit{the statistical description of turbulence}
and in particular for the so-called \textit{homogeneous, isotropic }and
\textit{stationary} turbulence (HIST). Here the problem is set in terms of
the $1-$point velocity probability density function (PDF) which determines a
complete IKT-statistical model for NS fluids. This raises the interesting
question of identifying the statistical assumptions under which a Gaussian
PDF can be achieved in such a context. In this paper it is proven that for
the IKT statistical model, HIST requires necessarily that $f_{1}$ must be
SIED (namely \textit{stationary}, \textit{isotropic} and \textit{%
everywhere-defined}). This implies, in turn, that the functional form of the
PDF is uniquely prescribed at all times. In particular, it is found that
necessarily the PDF must coincide with an isotropic Gaussian distribution.
The conclusion is relevant for the investigation of the so-called
homogenous, isotropic and stationary turbulence.
\end{abstract}

\date{\today }
\maketitle

\pagebreak

\section{Introduction: hydrodynamic turbulence}

The turbulence problem of hydrodynamics is made difficult by the fact that
there does not exist a definition of the turbulent state that is universally
accepted. The concept of fluid state is nevertheless well defined in fluid
dynamics, as it follows from the theory of continuum media. In particular,
it is prescribed by an appropriate set of suitably smooth real functions $%
\left\{ Z\right\} \equiv \left\{ Z_{i},i=1,n\right\} $ denoted as fluid
fields, which must be physically realizable, i.e., identified with physical
observables. The fluid fields by assumption satisfy a well-posed
initial-boundary value problem represented by a set of PDEs denoted as fluid
equations. In the theory of continua this means that they are necessarily
described by strong solutions (i.e., they are defined and are least
continuous everywhere in the existence domain). The same type of requirement
is also manifestly imposed by comparison with experimental observations.

Despite these premises, a precise mathematical definition of the concept of
"turbulent state" is still missing. In experiments on isolated
incompressible viscous fluids, turbulence is typically associated with the
manifestation of fluid motion in which the state of the fluid is
"turbulent", i.e., is in some sense random and decays in time until the
fluid comes ultimately to a state of rest. In the past at least three
different views have been adopted in this regard.

The first one is the \ mathematical viewpoint represented by the Leray
theory of turbulence (see Stewart,1988 \cite{Stewart1988}) according to
which "...turbulence is a fundamentally different problem from smooth flow",
which cannot be described in term of strong solutions. This conjecture
actually lead Leray (1931, \cite{Leray1933,Leray1934a,Leray 1934b}) to
introduce the concept of weak solutions, i.e., solutions which are not
defined, and are not continuous, everywhere in the existence domain.
Although appealing and extremely fruitful for its mathematical implications
the theory is manifestly un-physical and therefore should be rejected.

The second one is the so-called deterministic theory of turbulence, for
which turbulence should be produced, instead, by the occurrence of a strange
attractor (Ruelle-Takens theory; Berge, 1984 \cite{Berge1988}). According to
this view, there should exist a classical dynamical system, characterizing
in some sense the time evolution of the fluid fields, which should exhibit a
chaotic behavior on a suitable invariant (hyper-)surface. The basic
manifestation of chaos for such a dynamical system would be the occurrence
of an infinite cascade of quasi-periodic phenomena in which infinitely many
periods are sequentially generated in order to give the appearance of
randomness (Hopf-Landau theory; Landau 1959 \cite{Landau1959}).

However, the third, and most popular approach, is probably the statistical
theory of turbulence, which historically can be referred primarily to the
work of Kolmogorov (Kolmogorov, 1941 \cite{Kolmogorov1941}) and Hopf (Hopf,
1950/51 \cite{Hopf1950/51}). The statistical treatment of fluids usually
adopted for turbulent flows (which may be invoked, however, to describe also
regular flows) consists, instead, in the introduction of appropriate
axiomatic approaches denoted as \textit{statistical models }$\ \left\{
f,\Gamma \right\} $.

Their construction involves, besides the specification of the phase space ($%
\Gamma $) and the \textit{probability density function }(PDF) $f,$ the
identification of the functional class to which $f$\textit{\ }must belong,
denoted as $\left\{ f\right\} .$ As a consequence, the complete set of fluid
fields $\left\{ Z\right\} ,$ or only a proper subset as in the case of the
Monin-Lundgren approach \cite{Monin1967,Lundgren1967}, are expressed in
terms of suitable functionals (called \textit{moments}) of $f.$

\subsection{Stochastic representation of turbulence}

A basic aspect of the statistical description is the introduction of an
ensemble average operator acting on the possible realizations of the fluid.
A possible definition of such operator may be achieved by representing the
fluid fields in terms of hidden variables \cite%
{Tessarotto2008-4,Tessarotto2009-1}. By definition they denote a suitable
set of independent variables $\mathbf{\alpha }=\left\{ \alpha
_{i},i=1,k\right\} \in V_{\mathbf{\alpha }}\subseteq \mathbf{R}^{k},$ with $%
k\geq 1,$ which cannot be known deterministically, i.e., are not observable.
In the context of turbulence theory these variables are necessarily
stochastic. This means that they are characterized by a suitable \textit{%
stochastic probability density} $g$ defined on $V_{\mathbf{\alpha
}}$ (see definitions and related discussion in the Appendix,
Subsection B), while the ensemble average $\left\langle \cdot
\right\rangle $ can be identified
with the stochastic-averaging $\left\langle \cdot \right\rangle _{\mathbf{%
\alpha }}$ defined by Eq.(\ref{stochastic averaging operator}) [see Appendix
A]. \ Hence, for turbulent flows the fluid fields - together with the PDF $%
f_{1}$ and the vector field $\mathbf{F}(\mathbf{x},t,\mathbf{\alpha })$ -
can be assumed to admit a\textit{\ }representation of the form \cite%
{Tessarotto2008-4,Tessarotto2009-1}

\begin{equation}
\left\{
\begin{array}{c}
\left\{ Z\right\} =\left\{ Z(\mathbf{r},t,\mathbf{\alpha })\right\}  \\
f_{1}=f_{1}(\mathbf{r,u,}t,\mathbf{\alpha }) \\
\mathbf{F}=\mathbf{F}(\mathbf{x},t,\mathbf{\alpha })%
\end{array}%
\right.   \label{stochastic functions}
\end{equation}%
to be defined in terms of a set of \ hidden variables $\mathbf{\alpha }$ and
a stochastic model $\left\{ g,V_{\mathbf{\alpha }}\right\} $ (see again
Subsection B in the Appendix). Hence, $\left\{ Z\right\} ,$ $f_{1}$ and $%
\mathbf{F}(\mathbf{x},t,\mathbf{\alpha })$ are necessarily \textit{%
non-observable}. Nevertheless, if we assume that the fluid fields $\left\{
Z\right\} $ are uniquely-prescribed ordinary functions of $(\mathbf{x,}t,%
\mathbf{\alpha })$ defined for all $(\mathbf{x,}t,\mathbf{\alpha
})\in \overline{\Gamma }\times I\times V_{\alpha },$ it follows
that they can still be considered \textit{conditional observables
}(see Appendix, Subsection A). Similar conclusions apply to
$f_{1},$ and to the vector field
$\mathbf{F}(\mathbf{x},t,\mathbf{\alpha })$ as well$.$

\subsection{The Navier-Stokes dynamical system}

It is, however, generally agreed that a common important property should
characterize all NS fluids, either regular or turbulent: this is related to
the existence of the \textit{phase-space dynamical system}, to be denoted as
\textit{NS dynamical system}, which advances in time the complete set of
fluid fields defining the fluid state.

In other words, should such a dynamical system actually exist, it
would permit to cast the complete set of fluid equations in terms
of an equivalent (and possibly infinite) set of ordinary
differential equations which define the dynamical system itself.
For contemporary science the determination of such a dynamical
system represents not simply an intellectual challenge, but a
fundamental prerequisite for the proper formulation of all
phenomenological theories which are based on the description of
these fluids, and hence \textit{both for the deterministic and
statistical approaches to turbulence}. These involve, for example,
the understanding of the phase-space Lagrangian description of
fluids \cite{Tessarotto2008-3,Tessarotto2008-5} relevant to
determine tracer-particle dynamics
\cite{Tessarotto2008-6,Tessarotto2009b} as well as the
time-evolution of scalar and tensor fields in turbulent flows, the
search of exact(or approximate) kinetic closure conditions for
statistical models
(such as in the case of the Monin-Lundgren hierarchy \cite%
{Monin1967,Lundgren1967}), the investigation of stochastic models
\cite{Tessarotto2008-4,Tessarotto2008-7} able to reproduce
phenomenological data (such as the two-point velocity increments
PDFs \cite{Naert1997,Friedrich1999} ), the \textit{theoretical
prediction} of multi-point velocity probability densities, all
essential ingredients in fluid dynamics and in applied sciences.

Surprisingly, although phase-space descriptions of incompressible fluids
described by the incompressible NS equations (INSE) have been around for a
long time, starting from the historical work of Hopf (see also Hopf, 1952
\cite{Hopf1950/51}), Edwards (Edwards, 1964 \cite{Edwards1964}) and Rosen
(Rosen, 1971 \cite{Rosen1971}), until recently \cite{Tessarotto2004} the
problem [of the search of the NS dynamical system] has remained unsolved.
Its solution for the incompressible NS equations (see also Refs. \cite%
{Tessarotto2007,Tessarotto2009-1} for its extension to quantum and magneto
fluids) is based on the construction of a statistical model $\left\{
f_{1},\Gamma \right\} $ for the 1-point PDF $f_{1}$ which is required to
obey a Liouville equation and whose moments determine - via suitable
velocity-moments - the complete set of fluid fields which describe the state
of the fluid. \ As indicated elsewhere \cite{Tessarotto2008-4}, the approach
can be extended also to the statistical treatment of turbulence theory.

\section{Motivations}

An unsolved problem in the statistical theory of turbulence concerns the
so-called \textit{homogeneous, isotropic }and \textit{stationary} turbulence
(HIST) arising in incompressible Navier-Stokes (NS) fluids. This concerns in
particular the determination of the form of the $1-$point probability
density function (PDF), $f_{1},$ occurring in the presence of HIST, which
should uniquely determine, in turn, the statistical model $\left\{ f,\Gamma
\right\} $.

According to some authors (see in particular Batchelor \cite{Batchelor})
this is predicted as \textit{almost-Gaussian}, while others \cite%
{Falkovich1997,Li 2005} have pointed that the tails of the PDF might exhibit
a strongly \textit{non-Gaussian behavior}. Despite recent attempts at
possible a theoretical explanation \cite{Hosokawa2008}, still missing is a
definite answer to the question whether a generalized behavior of this type
should actually be expected or not. Apart insufficient experimental
evidence, a major difficulty is represented by the lack of a consistent
theoretical description of the PDF in the presence of HIST, permitting a
rigorous definite answer to this question. This raises the interesting
question whether, in some suitable setting, i.e., for appropriate
statistical models, the problem can actually be solved. Being a subject of
major importance - not only in fluid dynamics but also in statistical
mechanics and, as we intend to prove, in kinetic theory - the issue deserves
a careful investigation. The goal of this paper is to pose the problem in
the framework of the\textit{\ complete inverse kinetic theory} (IKT)
approach developed by Tessarotto \textit{et al.} \cite{Tessarotto2004} for
incompressible NS fluids (see in particular \cite%
{Tessarotto2008-4,Tessarotto2009-1}).

Here we intend to prove that \textit{based solely on IKT} the appropriate
form of the 1-point PDF in the presence of HIST can actually be \textit{%
uniquely} established for these fluids, based on suitable statistical
assumptions stemming from the requirement of existence of HIST and
appropriate initial conditions and the requirement that the initial 1-point
PDF is determined imposing PEM (principle of entropy maximization, Jaynes,
1957 \cite{Jaynes1957}; see also related discussion in Ref.\cite%
{Tessarotto2007}). This is shown to be described by a probability density $%
f_{1},$ defined on the\textit{\ restricted} phase-space $\Gamma =\Omega
\times U,$ with $\Omega $ and $U$ denoting respectively the configuration
space of the fluid [to be identified with a bounded subset of the Euclidean
space $%
%TCIMACRO{\U{211d} }%
%BeginExpansion
\mathbb{R}
%EndExpansion
^{3}$] and the Euclidean velocity space $U\subseteq
%TCIMACRO{\U{211d} }%
%BeginExpansion
\mathbb{R}
%EndExpansion
^{3},$ which fulfills the following properties (\#1-8):

\begin{enumerate}
\item (Property \#1) it depends explicitly on the fluid fields $\left\{ Z(%
\mathbf{r},t,\mathbf{\alpha })\right\} =\left\{ \mathbf{V}(\mathbf{r},t,%
\mathbf{\alpha }),p_{1}(\mathbf{r},t,\mathbf{\alpha })\right\} ,$ namely is
of the type $f_{1}=f_{1}(\mathbf{r,v,}t,\mathbf{\alpha }),$ with $\mathbf{V}(%
\mathbf{r},t,\mathbf{\alpha })$ denoting the fluid velocity and $p_{1}(%
\mathbf{r},t,\mathbf{\alpha })$ the kinetic pressure. In particular $p_{1}(%
\mathbf{r},t,\mathbf{\alpha })$ is defined as the strictly positive function
$p_{1}(\mathbf{r},t,\mathbf{\alpha })=p(\mathbf{r},t,\mathbf{\alpha }%
)+p_{0}(t,\mathbf{\alpha })+\phi (\mathbf{r},t,\mathbf{\alpha }),$ with $p(%
\mathbf{r},t,\mathbf{\alpha }),p_{0}(t,\mathbf{\alpha })$ and $\phi (\mathbf{%
r},t,\mathbf{\alpha })$ representing respectively the fluid pressure, the
(strictly-positive) pseudo-pressure and the (possible) potential associated
to the conservative volume force density acting on the fluid;

\item (Property \#2) $f_{1}$ is a \textit{velocity probability density, i.e.,%
} it is normalized in velocity space so that%
\begin{equation}
\int\limits_{U}d^{3}\mathbf{v}f_{1}(\mathbf{r,v,}t,\mathbf{\alpha })=1;
\label{NORMALIZATION}
\end{equation}

\item (Property \#3) $f_{1}$ is \textit{strictly positive} in the velocity
space $U;$

\item (Property \#4) $f_{1}$ \textit{Galilei invariant in velocity space},
namely it is invariant with respect to a transformation of the form:%
\begin{equation}
\left\{
\begin{array}{c}
\mathbf{v} \\
\mathbf{V}(\mathbf{r},t,\mathbf{\alpha })%
\end{array}%
\right. \rightarrow \left\{
\begin{array}{c}
\mathbf{v+V}_{o} \\
\mathbf{V}(\mathbf{r},t,\mathbf{\alpha })+\mathbf{V}_{o}%
\end{array}%
\right.   \label{TRANSLATION}
\end{equation}%
with $\mathbf{V}_{o}$ such that $\mathbf{v+V}_{o}\in U.$ As a consequence, $%
f_{1}$ is of the form%
\begin{equation}
f_{1}=f_{1}(\mathbf{r,u},t,\mathbf{\alpha }),  \label{HOMOGENEOUS PDF}
\end{equation}%
with $\mathbf{u\equiv u}(\mathbf{r},t,\alpha )\mathbf{=}\mathbf{\mathbf{v}}-%
\mathbf{V}(\mathbf{r},t,\mathbf{\alpha })$ denoting the relative velocity,
namely is \textit{homogeneous }in the velocity space $U$\textit{;}

\item (Property \#5) the velocity space $U$ coincides with $%
%TCIMACRO{\U{211d} }%
%BeginExpansion
\mathbb{R}
%EndExpansion
^{3}.$ $f_{1}$ fulfilling this property is said \textit{everywhere defined }%
(in\textit{\ }$U\equiv
%TCIMACRO{\U{211d} }%
%BeginExpansion
\mathbb{R}
%EndExpansion
^{3}$)$.$

\item (Property \#6) $f_{1}$ is \textit{stationary,} namely it can depend on
time only via the fluid fields:%
\begin{equation}
f_{1}=f_{1}(\mathbf{r,u},\mathbf{\alpha }),  \label{STATIONARY}
\end{equation}

\item (Property \#7) $f_{1}$ is \textit{isotropic} in velocity space, i.e.,
it is of the form
\begin{equation}
f_{1}=f_{1}(\mathbf{r,}\left\vert \mathbf{u}\right\vert ,\mathbf{\alpha });
\label{ISOTROPIC}
\end{equation}

\item (Property \#8) $f_{1}$ is a\textit{\ Gaussian distribution} of the form%
\begin{equation}
f_{M}(\mathbf{r,}\left\vert \mathbf{u}\right\vert \mathbf{,}p_{1}(\mathbf{%
\mathbf{r}},t\mathbf{,\mathbf{\alpha }}))=\frac{1}{\pi ^{3/2}v_{thp}^{3}(%
\mathbf{r,}t,\mathbf{\alpha })}\exp \left\{ -\frac{u^{2}}{v_{thp}^{2}(%
\mathbf{r,}t,\mathbf{\alpha })}\right\} ,  \label{GAUSSIAN PDF}
\end{equation}%
with $v_{thp}(\mathbf{r,}t,\mathbf{\alpha })=\sqrt{2p_{1}(\mathbf{\mathbf{r}}%
,t,\alpha )/\rho _{o}}$ denoting the thermal velocity associated to the
kinetic pressure.
\end{enumerate}

\section{Homogeneous, isotropic and stationary turbulence}

In fluid dynamics two types of descriptions of the fluid, respectively
denoted \textit{deterministic} and \textit{stochastic}, can be
distinguished, in which the fluid fields describing the state of the fluid
are treated respectively as deterministic or stochastic functions. In both
cases the fluid fields $\left\{ Z\right\} \equiv \left\{ Z_{i},i=1,n\right\}
$ are considered, in a suitable existence domain, suitably smooth strong
solutions of the fluid equations. In the so-called \textit{statistical
theory of turbulence}, historically referred primarily to the work of
Kolmogorov (Kolmogorov, 1941 \cite{Kolmogorov1941}) and Hopf (Hopf, 1950/51
\cite{Hopf1950/51}), turbulence is intended as the characteristic property
of the fluid in which the fluid fields $\left\{ Z\right\} $ can only be
prescribed in a statistical sense. This implies that they are necessarily
stochastic functions of the form $\left\{ Z\right\} =\left\{ Z(\mathbf{r},t,%
\mathbf{\alpha })\right\} $ characterized by \textit{a stochastic PDF} $g(%
\mathbf{r,}t,\mathbf{\alpha })$ defined for all $\left( \mathbf{r,}t\right)
\in \overline{\Omega }\times I$ \cite{Tessarotto2008-4,Tessarotto2009-1} and
$\mathbf{\alpha }=\left\{ \alpha _{i},i=1,k\right\} \in V_{\mathbf{\alpha }%
}\subseteq \mathbf{R}^{k}$ (with $k\geq 1$), with $\mathbf{\alpha }$
suitable stochastic variables independent of\textit{\ }$\mathbf{r,}t$. Thus,
introducing the \textit{stochastic-averaging operator} $\left\langle \cdot
\right\rangle _{\mathbf{\alpha }}\equiv \int\limits_{V_{\mathbf{\alpha }}}d%
\mathbf{\alpha }g(\mathbf{r,}t,\mathbf{\alpha })\cdot ,$ acting on an
arbitrary integrable function it follows that the fluid fields $Z_{i}(%
\mathbf{r,}t,\mathbf{\alpha })$ (for $i=1,n$) can always be represented in
terms of their stochastic decompositions $Z_{i}=\left\langle
Z_{i}\right\rangle _{\mathbf{\alpha }}+\delta Z_{i},$ with $\left\langle
Z_{i}\right\rangle _{\mathbf{\alpha }}$ denoting their stochastic averages.

A widespread conjecture is that turbulence, at least in special
circumstances [usually ascribed to the so-called "fully developed"
turbulence (FDT)], should be characterized by certain \textit{universal
properties}. These concern, in particular, the concept of homogeneous,
isotropic and stationary turbulence (HIST). Its definition \cite%
{Kolmogorov1941} (see also Refs.\cite{Monin1975,Frisch1995}) is related to
the assumed properties of the operator $\left\langle \cdot .\right\rangle _{%
\mathbf{\alpha }}$ and of the velocity increments%
\begin{equation}
dV_{i}\equiv V_{i}(\mathbf{r}_{1},t)-V_{i}(\mathbf{r},t),
\end{equation}%
which are assumed to be defined for arbitrary displacements
\begin{equation}
\mathbf{dr=r}_{1}-\mathbf{r}  \label{diplacement}
\end{equation}%
such that both $\mathbf{r}$ and $\mathbf{r}_{1}$ belong to $\overline{\Omega
}$.

\subsubsection{Definition - HIST}

Turbulence is said \textit{homogeneous, isotropic and stationary} if:

\begin{itemize}
\item \textit{HIST Requirement \#1:} the stochastic-averaging operator $%
\left\langle Z_{i}\right\rangle _{\mathbf{\alpha }}$ commutes with all the
differential operators appearing in the fluid equations (namely, for the NS
equations, this means it must commute with the operators $\frac{\partial }{%
\partial t},\nabla $ and $\nabla ^{2}$);

\item \textit{HIST Requirement \#2:} for all $n\in \mathbb{N}_{0}$ \ the
\textit{structure functions }- i.e., the stochastic-averages of $\left\{
dV_{i}\right\} ^{n},$ with $S_{i}^{(n)}(\mathbf{r,dr},t)\equiv \left\langle
\left\{ dV_{i}\right\} ^{n}\right\rangle _{\alpha }$ - are respectively:

2$_{a}$) \textit{\ }independent of $\mathbf{r,}$ namely for all $i=1,2,3$%
\begin{equation}
S_{i}^{(n)}=S_{i}^{(n)}(\mathbf{dr,}t)
\end{equation}%
(\textit{homogeneous turbulence}) $;$

2$_{b}$) independent of the directions of $\mathbf{dr}$ and $\mathbf{V}$,
hence for all $i=1,2,3:$%
\begin{equation}
S_{i}^{(n)}=S^{(n)}(\mathbf{r,}l\mathbf{,}t)
\end{equation}%
(\textit{isotropic turbulence}), where $l=\left\vert \mathbf{dr}\right\vert $
is the magnitude of the displacement (\ref{diplacement})$;$

2$_{c}$)independent of $t$, hence for all $i=1,2,3:$%
\begin{equation}
S_{i}^{(n)}=S_{i}^{(n)}(\mathbf{r,dr})
\end{equation}
(\textit{stationary turbulence}).
\end{itemize}

Thus, for all $n\in \mathbb{N}_{0}$ and $i=1,2,3$,\ HIST is by assumption
characterized by structure functions of the form
\begin{equation}
S_{i}^{(n)}=S^{(n)}(l),  \label{REQUIREMENT ON STRUCTURE
FUNCTIONS}
\end{equation}%
i.e., depending solely on the magnitude of the displacement (\ref%
{diplacement}).

\section{The IKT-statistical model for turbulent fluids}

Starting point for the statistical treatment of turbulence in NS fluids in
the IKT approach is the introduction of a statistical model $\left\{
f_{1},\Gamma \right\} $ (\textit{to be denoted as IKT-statistical model})
for the INSE problem. The corresponding fluid fields\ are $\left\{ \rho _{o},%
\mathbf{V}(\mathbf{r},t,\mathbf{\alpha }),p(\mathbf{r},t,\mathbf{\alpha }%
),S_{T}\right\} $ with $\rho _{o}$ and $S_{T}$ to be identified,
respectively, with the constant mass density and the constant thermodynamic
entropy$.$ In the following the fluid fields are required to be: 1) they are
strong solutions of the INSE problem in $\overline{\Omega }\times I\times V_{%
\mathbf{\alpha }},$ with bounded configuration space $\overline{\Omega }$ (%
\textit{internal domain}) ; 2) global solutions, i.e., defined for all $t\in
I\equiv
%TCIMACRO{\U{211d} }%
%BeginExpansion
\mathbb{R}
%EndExpansion
.$

The construction of $\left\{ f_{1},\Gamma \right\} $ \cite{Tessarotto2008-4}
involves the definition of a suitable PDF $f_{1}$ defined on a phase space $%
\Gamma $, \textit{denoted as }$1$\textit{-point velocity PDF}, which permits
the representation, via a suitable mapping $\left\{ f_{1},\Gamma \right\}
\Rightarrow \left\{ Z\right\} ,$ $\left\{ Z\right\} $ denoting \textit{the
complete set }of the \textit{fluid fields} $\left\{ Z\right\} \equiv \left\{
Z_{i},i=1,n\right\} $ defining the state of the fluid. In particular by
assumption $\Gamma $ is identified with the restricted phase-space $\Gamma
=\Omega \times U$ $\times V_{\mathbf{\alpha }}$ [with closure $\overline{%
\Gamma }=\overline{\Omega }\times U\times V_{\mathbf{\alpha }}$];
furthermore, the $1$-point velocity PDF:

\begin{enumerate}
\item is taken of the general form $f_{1}(t)\equiv f_{1}(\mathbf{x},t,%
\mathbf{\alpha }),$ with $\mathbf{x}=(\mathbf{r,v}),$ where $\mathbf{r}\in $
$\overline{\Omega }$ and $\mathbf{v}\in U\subseteq
%TCIMACRO{\U{211d} }%
%BeginExpansion
\mathbb{R}
%EndExpansion
^{3}$ (with $U$ defined as the open subset of $%
%TCIMACRO{\U{211d} }%
%BeginExpansion
\mathbb{R}
%EndExpansion
^{3}$ spanned by $\mathbf{v}$ on which $f_{1}>0$)$.$ In addition $f_{1}$ is
by assumption Galilei invariant and hence invariant w.r. to (\ref%
{TRANSLATION}). It follows that $f_{1}$ is necessarily homogeneous in
velocity space, namely of the form (\ref{HOMOGENEOUS PDF});

\item determines in terms of suitable moments \textit{the complete set of
the fluid fields }$\left\{ Z\right\} $ which define the state of the fluid.
This requires that the fluid fields are determined by the velocity moments
\begin{equation}
\int\limits_{U}d^{3}\mathbf{v}G(\mathbf{x},t,\mathbf{\alpha })f_{1}(\mathbf{x%
},t,\mathbf{\alpha })=\mathbf{V}(\mathbf{r},t,\mathbf{\alpha }),p_{1}(%
\mathbf{r},t,\mathbf{\alpha }),  \label{MOMENTS-1}
\end{equation}%
defined respectively for the weight-functions $G(\mathbf{x},t,\mathbf{\alpha
})=\mathbf{v,}\rho _{o}u^{2}/3,$ whereas $S_{T}$ is identified with the
Boltzmann-Shannon entropy, i.e., the phase-space moment
\begin{equation}
S(f_{1}(t))\equiv -\int\limits_{\Gamma }d\mathbf{x}f_{1}(\mathbf{x},t,%
\mathbf{\alpha })\ln f_{1}(\mathbf{x},t,\mathbf{\alpha }),
\label{B-S entropu}
\end{equation}%
by imposing that $\forall t\in I\equiv
%TCIMACRO{\U{211d} }%
%BeginExpansion
\mathbb{R}
%EndExpansion
$ the constraint \cite{Tessarotto2007}
\begin{equation}
S_{T}=S(f_{1}(t)).  \label{constant entropy}
\end{equation}
\end{enumerate}

The set of equations (\ref{MOMENTS-1}) and (\ref{B-S entropu}), denoted as
\textit{correspondence principle,} are assumed do be fulfilled identically
in the whole existence domain of the fluid fields $\mathbf{V}(\mathbf{r},t,%
\mathbf{\alpha }),p_{1}(\mathbf{r},t,\mathbf{\alpha })$ and $S_{T}$. This
implies manifestly that $f_{1}(\mathbf{x},t,\mathbf{\alpha })$ must be
defined and strictly positive on $\overline{\Gamma }=\overline{\Omega }%
\times U$ $\times V_{\mathbf{\alpha }}$ . The time-evolution of $f_{1}(%
\mathbf{x},t,\mathbf{\alpha })$ is then uniquely determined by the flow {(%
\textit{stochastic N-S dynamical system})}
\begin{equation}
T_{t_{o},t}:\mathbf{x}_{o}\rightarrow \mathbf{x}(t)=T_{t_{o},t}\mathbf{x}%
_{o},  \label{NS DS}
\end{equation}%
with $T_{t_{o},t}$ \ the corresponding evolution operator generated by the
initial value problem{%
\begin{eqnarray}
&&\left. \frac{d}{dt}\mathbf{x}=\mathbf{X}(\mathbf{x},t,\mathbf{\alpha }%
)\right.  \\
&&\left. \mathbf{x(}t_{o})=\mathbf{x}_{o}.\right.   \notag
\end{eqnarray}%
In particular, denoting by }$J(t)=\left\vert \frac{\partial \mathbf{x}(t)}{%
\partial \mathbf{x}_{o}}\right\vert =\exp \left\{
\int\limits_{t_{o}}^{r}dt^{\prime }\frac{\partial }{\partial \mathbf{x}%
(t^{\prime })}\cdot \mathbf{X}(\mathbf{x}(t^{\prime }),t^{\prime },\mathbf{%
\alpha })\right\} $ the Jacobian determinant of the flow the time-advanced
PDF $f_{1}(\mathbf{x}(t),t,\mathbf{\alpha })$ is uniquely determined
requiring that for all $\left( \mathbf{x}_{o},\mathbf{\alpha ,}%
t_{o},t,\right) $ $\in \overline{\Gamma }\times I\times I$ it satisfies the
\emph{Lagrangian} \emph{inverse kinetic equation }(IKE)%
\begin{equation}
f_{1}(t)=J(t)f_{1}(t_{o}).  \label{LAGRANGIAN IKE}
\end{equation}%
In particular, as shown elsewhere \cite{Tessarotto2009-1}, the initial PDF $%
f_{1}(t_{o})\equiv f(\mathbf{x,}t_{o},\mathbf{\alpha })$ can be assumed to
be a strictly positive smooth PDF of the form
\begin{equation}
f_{1}(t_{o})=\left\langle f_{1}(t_{o})\right\rangle _{\Omega }\frac{h(t_{o})%
}{\left\langle h(t_{o})\right\rangle _{\Omega }},  \label{position}
\end{equation}%
with $h(t_{o})\equiv h(\mathbf{x}_{o},t_{o},\mathbf{\alpha })$ to be
determined and $\left\langle f_{1}(t_{o})\right\rangle _{\Omega }$ subject
to the constraint
\begin{equation}
\left\langle f_{1}(t_{o})\right\rangle _{\Omega }=\widehat{f}%
_{1}^{(freq)}(t_{o},\mathbf{v}_{o},\mathbf{\alpha })  \label{position-2}
\end{equation}%
(\textit{physical realizability condition}). Here $\widehat{f}%
_{1}^{(freq)}(t_{o},\mathbf{v}_{o},\mathbf{\alpha })$ denotes a suitable
\textit{continuous velocity-frequency density, }uniquely associated to the
initial fluid velocity\textit{\ }$\mathbf{V}(\mathbf{r}_{o},t_{o},\mathbf{%
\alpha }).$ Furthermore, $\left\langle f_{1}(t)\right\rangle _{\Omega }$
denotes the configuration-space average (defined at at time $t$), i.e.,
\begin{equation}
\left\langle f_{1}(\mathbf{r,v},t,\mathbf{\alpha })\right\rangle _{\Omega
}\equiv \frac{1}{\mu (\Omega )}\int\limits_{\Omega }d^{3}\mathbf{r}f_{1}(%
\mathbf{r,v},t,\mathbf{\alpha }),
\end{equation}%
while $\mu (\Omega )=\int\limits_{\Omega }d^{3}\mathbf{r}>0$ is the
canonical measure of $\Omega .$ As a consequence $h(t_{o}),$ can be uniquely
determined imposing that $f_{1}(\mathbf{x}_{o},t_{o},\mathbf{\alpha })$
satisfies PEM \cite{Jaynes1957}, namely the variational principle $\delta
S(f_{1}(t))=0$ subject to the constrains (\ref{position}) and (\ref%
{position-2}), taking the form of a \emph{generalized Gaussian distribution\
}
\begin{equation}
h(t_{o})=\exp \left\{ -1-\lambda _{o}(\mathbf{r}_{o}\mathbf{,}t_{o},\mathbf{%
\alpha })-\lambda _{2i}(\mathbf{r}_{o}\mathbf{,}t_{o},\mathbf{\alpha }%
)u_{i}^{2}(\mathbf{r}_{o}\mathbf{,}t_{o},\mathbf{\alpha })\right\} .
\label{NON-ISOTROPIX GAUSSIAN}
\end{equation}%
Here, $\lambda _{o}(\mathbf{r}_{o}\mathbf{,}t_{o},\mathbf{\alpha })$\textit{%
\ and}\textbf{\ }$\lambda _{2i}(\mathbf{r}_{o}\mathbf{,}t_{o},\mathbf{\alpha
})$ (for $i=1,2,3$) denote suitable Lagrange multipliers to be determined
imposing the moment equations (\ref{MOMENTS-1}) and (\ref{B-S entropu}),
together with the constraint (\ref{constant entropy}). Furthermore, $h(t_{o})
$ can be shown to take the form of an \emph{isotropic Gaussian distribution}%
, i.e., \emph{\ }%
\begin{equation}
h_{o}(t)=\exp \left\{ -1-\lambda _{o}(\mathbf{r,}t,\mathbf{\alpha })-\lambda
_{2}(\mathbf{r,}t,\mathbf{\alpha })u^{2}(\mathbf{r,}t,\mathbf{\alpha }%
)\right\}   \label{Gaussian distribution}
\end{equation}%
[see related discussion the Appendix].

The equivalence theorem pointed out elsewhere \cite{Tessarotto2010c} between
the INSE problem and the NS dynamical system (\ref{NS DS}) then warrants the
validity of the correspondence principle, i.e., that the moments prescribed
by equations (\ref{MOMENTS-1}) and (\ref{B-S entropu}) actually define a
strong solution of the INSE problem.

\section{Connection with HIST - SIED IKT-statistical models}

Here we are interested in determining a particular subclass of
IKT-statistical models $\left\{ f_{1},\Gamma \right\} $ which may fulfill at
least some of the properties which characterize HIST for NS fluids. Let us
now analyze the consequences placed by (the assumption) of HIST. First, let
us require that - consistent with the requirement of FDT - the velocity
space $U$ [on which $f_{1}$ is defined and strictly positive] \textit{%
coincides with} $\mathbb{R}^{3}$ and hence $f_{1}$ is \textit{everywhere
defined}. Second, we notice that the constraints imposed by Eq.(\ref%
{REQUIREMENT ON STRUCTURE FUNCTIONS}) on arbitrary structure functions $%
S_{i}^{(n)}$ can generally be satisfied only if:

\begin{itemize}
\item $f_{1}(t)$ if stationary in the sense (\ref{STATIONARY});

\item $f_{1}(t)$ isotropic in velocity space (since no preferred direction
in velocity space can exist in such a case).
\end{itemize}

As a consequence, in the presence of HIST, $f_{1}$ is necessarily \textit{%
stationary, isotropic} and\textit{\ everywhere defined} (SIED). In the
following we shall denote as SIED the IKT-statistical models $\left\{
f_{1},\Gamma \right\} $ which fulfill these requirements. In such a case it
is immediate to reach the following result [which proves Properties \#1-8]:

\noindent \textbf{THM.1 - Characteristic property of SIED $\left\{
f_{1},\Gamma \right\} $}

\emph{For SIED IKT-statistical models }$\left\{ f_{1},\Gamma \right\} $
\emph{there results identically in }$\overline{\Gamma }\times I:$

\begin{eqnarray}
h_{o}(t) &=&f_{M}(\mathbf{r,}\left\vert \mathbf{u}(\mathbf{\mathbf{r},}t%
\mathbf{,\mathbf{\alpha }})\right\vert \mathbf{,}p_{1}(\mathbf{\mathbf{r}},t%
\mathbf{,\mathbf{\alpha }})),  \label{THESIS -1} \\
\frac{\left\langle f_{1}(t)\right\rangle _{\Omega }}{\left\langle
h(t)\right\rangle _{\Omega }} &=&1,  \label{THESIS -2}
\end{eqnarray}%
\emph{where }$f_{M}(\mathbf{r,}\left\vert \mathbf{u}(\mathbf{\mathbf{r},}t%
\mathbf{,\mathbf{\alpha }})\right\vert \mathbf{,}p_{1}(\mathbf{\mathbf{r}},t%
\mathbf{,\mathbf{\alpha }}))$ \emph{is the Gaussian 1-point PDF (\ref%
{GAUSSIAN PDF}) and }$h_{o}(t)$ \emph{the Gaussian distribution (\ref%
{Gaussian distribution}). In particular,} $\lambda (\mathbf{r,}t,\mathbf{%
\alpha })$\textit{\ }\emph{and\ }$\lambda _{2}(\mathbf{r,}t,\mathbf{\alpha }%
) $ \emph{(for }$i=1,2,3$\emph{) denote suitable Lagrange multipliers to be
determined imposing the moment equations (\ref{MOMENTS-1}) and (\ref{B-S
entropu}), together with the constraint (\ref{constant entropy}).}

\noindent PROOF In fact, first, since $f_{1}(t)$ is stationary for all $t\in
I\equiv \mathbb{R},$ $f_{1}(t)$ is necessarily determined by PEM ,subject to
the constrains (\ref{position}) and (\ref{position-2}). Therefore for all $%
t\in I\equiv \mathbb{R},$ $f_{1}(t)$ has the form
\begin{equation}
f_{1}(t)=\left\langle f_{1}(t)\right\rangle _{\Omega }\frac{h(t)}{%
\left\langle h(t)\right\rangle _{\Omega }},
\end{equation}%
with $h(t)\equiv h(\mathbf{r,v,}t,\mathbf{\alpha })$ of the type (\ref%
{NON-ISOTROPIX GAUSSIAN}). Furthermore, since $\left\{ f_{1},\Gamma \right\}
$ is by assumption isotropic and everywhere-defined (so that $U\equiv
\mathbb{R}^{3}$) it follows necessarily for all $(\mathbf{v,t})\in $ $%
\mathbb{R}^{3}\times I$ that%
\begin{eqnarray}
&&\left. \frac{\left\langle f_{1}(t)\right\rangle _{\Omega }}{\left\langle
h(t)\right\rangle _{\Omega }}=c(t,\mathbf{\alpha })>0,\right.  \\
&&\left. h(t)=\exp \left\{ -1-\lambda _{o}(\mathbf{r,}t,\mathbf{\alpha }%
)-\lambda _{2}(\mathbf{r,}t,\mathbf{\alpha })u^{2}\right\} \equiv
h_{o}(t).\right.
\end{eqnarray}%
Here $c(t,\mathbf{\alpha }),$ $\lambda _{o}(\mathbf{r,}t,\mathbf{\alpha })$
and $\lambda _{2}(\mathbf{r,}t,\mathbf{\alpha })$ denote, respectively, a
suitable strictly positive function of time and two Lagrange multipliers
defined so that there results identically in $\overline{\Omega }\times I$
\begin{eqnarray}
&&\left. \int\limits_{\mathbb{R}^{3}}d^{3}vf_{1}(t)=1,\right.  \\
&&\left. \int\limits_{\mathbb{R}^{3}}d^{3}v\frac{\rho _{o}u^{2}}{3}%
f_{1}(t)=p_{1}(\mathbf{r},t,\mathbf{\alpha }),\right.
\end{eqnarray}%
while the pseudo-pressure $p_{0}(t,\mathbf{\alpha })$ must be determined by
imposing the constraint (\ref{constant entropy}). \ Hence, it is always
possible to set $c(t,\mathbf{\alpha })=1,$ so that (\ref{THESIS -2}) is
identically satisfied. As a consequence, it follows identically that $%
h_{o}(t)\equiv f_{M}(\mathbf{r,}\left\vert \mathbf{u}(\mathbf{\mathbf{r},}t%
\mathbf{,\mathbf{\alpha }})\right\vert \mathbf{,}p_{1}(\mathbf{\mathbf{r}},t%
\mathbf{,\mathbf{\alpha }})),$ which proves Eq.(\ref{THESIS -1}) too. Q.E.D.

\section{Conclusions}

In this paper properties of the IKT-statistical model, $\left\{ f_{1},\Gamma
\right\} $, defined in terms of the 1-point PDF $f_{1}$ for a turbulent
fluid obeying the INSE problem, have been investigated.

In particular we have proven that for the IKT statistical model $\left\{
f_{1},\Gamma \right\} $ :

\begin{itemize}
\item the assumption of HIST requires that $f_{1}$ must necessarily be SIED,
i.e., stationary and isotropic and everywhere-defined in velocity space;

\item the requirement of stationarity implies that PEM must hold identically
for all $t\in I\equiv \mathbb{R}.$ As a consequence the functional form of
the function $h(t)$ remains uniquely determined.
\end{itemize}

Main result is the proof here achieved (THM.1) that the requirement of $%
f_{1} $ to be SIED \textit{implies necessarily that }$f_{1}$\textit{\ must
coincides identically with an isotropic Gaussian distribution}.

The conclusion is relevant for the investigation of the so-called
homogenous, isotropic and stationary turbulence.

\section{Acknowledgments}

Work developed in cooperation with the CMFD Team, Consortium for
Magneto-fluid-dynamics (Trieste University, Trieste, Italy). Research
partially performed in the framework of the GDRE (Groupe de Recherche
Europeenne) GAMAS.

\section{Appendix: The mathematical description of incompressible NS fluids}

In fluid dynamics the state of an arbitrary fluid system is assumed to be
defined everywhere in a suitable \textit{extended configuration domain} $%
\Omega \times I$ [$\Omega $ \ denoting the configuration space and $%
I\subseteq
%TCIMACRO{\U{211d} }%
%BeginExpansion
\mathbb{R}
%EndExpansion
$ the time axis] by an appropriate set of suitably smooth functions $\left\{
Z\right\} ,$ denoted as \textit{fluid fields, }and by a well-posed set of
PDEs, denoted as \textit{fluid equations, }of which the former are\textit{\ }%
solutions\textit{.} The fluid fields are by assumption functions of the
observables $(\mathbf{r,}t)$, with $\mathbf{r}$ and $t$ spanning
respectively the sets $\Omega $ and $I,$ namely smooth real functions.
Therefore, they are also \textit{strong solutions} of the fluid equations.
In particular, this means that they are are required to be at least
continuous in all points of the closed set $\overline{\Omega }\times I$,
with $\overline{\Omega }=\Omega \cup \partial \Omega $ closure of $\Omega ${$%
.$ \ In the remainder we shall require, for definiteness, that:}

\begin{enumerate}
\item {$\Omega ${\ (\textit{configuration domain}) is a bounded subset of
the Euclidean space }$E^{3}$ on }$%
%TCIMACRO{\U{211d} }%
%BeginExpansion
\mathbb{R}
%EndExpansion
${$^{3};$}

\item $I$ (\textit{time axis}) is identified, when appropriate, either with
a bounded interval, \textit{i.e.}, $I${$=$}$\left] {t_{0},t_{1}}\right[
\subseteq
%TCIMACRO{\U{211d} }%
%BeginExpansion
\mathbb{R}
%EndExpansion
,$ or with the real axis $%
%TCIMACRO{\U{211d} }%
%BeginExpansion
\mathbb{R}
%EndExpansion
$;

\item in the open set $\Omega \times ${$I$} the functions $\left\{ Z\right\}
,$ are assumed to be solutions of a closed set of fluid equations\textit{. }%
In the case of an incompressible Navier-Stokes fluid {t}he fluid fields are $%
\left\{ Z\right\} \mathbf{\equiv }\left\{ \mathbf{V},p,S_{T}\right\} $ and
their fluid equations
\begin{eqnarray}
\rho &=&\rho _{o},  \label{1b} \\
\nabla \cdot \mathbf{V} &=&0,  \label{1ba} \\
N\mathbf{V} &=&0,  \label{1bbb} \\
\frac{\partial }{\partial t}S_{T} &=&0,  \label{1bb} \\
Z(\mathbf{r,}t_{o}) &\mathbf{=}&Z_{o}(\mathbf{r}),  \label{1ca} \\
\left. Z(\mathbf{r,}t)\right\vert _{\partial \Omega } &\mathbf{=}&\left.
Z_{w}(\mathbf{r,}t)\right\vert _{\partial \Omega },  \label{1c}
\end{eqnarray}%
where Eqs.(\ref{1b})-(\ref{1bb}) denote the{\ \textit{incompressible
Navier-Stokes equations} (INSE) and Eqs. (\ref{1b})- (\ref{1c}) the
corresponding \textit{initial-boundary value INSE problem}. In particular},
Eqs. (\ref{1b})- (\ref{1c}) are respectively the \textit{incompressibility,
isochoricity, Navier-Stokes and constant thermodynamic entropy equations}
and the initial and Dirichlet boundary conditions for $\left\{ Z\right\} ,$
with $\left\{ Z_{o}(\mathbf{r})\right\} $ and $\left\{ \left. Z_{w}(\mathbf{%
r,}t)\right\vert _{\partial \Omega }\right\} $ suitably prescribed initial
and boundary-value fluid fields, defined respectively at the initial time $%
t=t_{o}$ and on the boundary $\partial \Omega .$

\item by assumption, these equations together with appropriate initial and
boundary conditions are required to define a well-posed problem with unique
strong solution defined everywhere in $\Omega \times ${$I$}.
\end{enumerate}

Here the notation as follows. $N$ is the \textit{NS nonlinear operator}
\begin{equation}
N\mathbf{V}=\frac{D}{Dt}\mathbf{V}-\mathbf{F}_{H},  \label{NS operator}
\end{equation}%
with $\frac{D}{Dt}\mathbf{V}$ and\textbf{\ }$\mathbf{F}_{H}$ denoting
respectively the \textit{Lagrangian fluid acceleration} and the \textit{%
total force} \textit{per unit mass}
\begin{eqnarray}
&&\left. \frac{D}{Dt}\mathbf{V}=\frac{\partial }{\partial t}\mathbf{V}+%
\mathbf{V}\cdot \nabla \mathbf{V,}\right. \\
&&\left. \mathbf{F}_{H}\equiv \mathbf{-}\frac{1}{\rho _{o}}\nabla p+\frac{1}{%
\rho _{o}}\mathbf{f}+\upsilon \nabla ^{2}\mathbf{V,}\right.  \label{2c}
\end{eqnarray}%
while $\rho _{o}>0$ and $\nu >0$ are the \textit{constant mass density} and
the constant \textit{kinematic viscosity}. In particular, $\mathbf{f}$ is
the \textit{volume force density} acting on the fluid, namely which is
assumed of the form%
\begin{equation}
\mathbf{f=-\nabla }\phi (\mathbf{r},t)+\mathbf{f}_{R},
\end{equation}%
$\phi (\mathbf{r},t)$ being a suitable scalar potential, so that the first
two force terms [in Eq.(\ref{2c})] can be represented as $-\nabla p+\mathbf{f%
}$ $=-\nabla p_{r}+\mathbf{f}_{R},$ with
\begin{equation}
p_{r}(\mathbf{r},t)=p(\mathbf{r},t)-\phi (\mathbf{r},t),
\end{equation}%
denoting the \textit{reduced fluid pressure}. As a consequence of Eqs.(\ref%
{1b}),(\ref{1ba}) and (\ref{1bbb}) it follows that the fluid pressure
necessarily satisfies the \textit{Poisson equation}%
\begin{equation}
\nabla ^{2}p=S,  \label{Poisson}
\end{equation}%
where the source term $S$ reads
\begin{equation}
S=-\rho _{o}\nabla \cdot \left( \mathbf{V}\cdot \nabla \mathbf{V}\right)
+\nabla \cdot \mathbf{f}.
\end{equation}

\subsection{Physical/conditional observables - Hidden variables}

The fluid fields $\left\{ Z\right\} $ are, by assumption, prescribed smooth
real functions of \ $(\mathbf{r},t)\in $ $\Omega \times I$. \ In particular,
they can\ be either \textit{physical observables}\textbf{\ }or \textit{%
conditional observable, }according to the definitions indicated below\textit{%
.}

\subsubsection{\textbf{Definition - }Physical observable\textbf{\textit{/}%
conditional observable}}

A \textit{physical observable} is an arbitrary real-valued and
uniquely-defined smooth real function of $(\mathbf{r,}t)\in $ $\Omega \times
I$. Hence, as a particular case $(\mathbf{r,}t)$ are observable too.

A \textit{conditional observable} is, instead, an arbitrary real-valued and
uniquely-defined smooth real function of $(\mathbf{r,}t)\in $ $\Omega \times
I$ which depends also on non-observable variables and is, as such, an
uniquely-prescribed function of the latter ones.

Therefore the functions $Z_{i}$ can be assumed respectively of the form \cite%
{Tessarotto2008-4,Tessarotto2009-1}%
\begin{equation}
Z_{i}\equiv Z_{i}(\mathbf{r},t)  \label{determistic fluid fields}
\end{equation}%
or
\begin{equation}
Z_{i}\equiv Z_{i}(\mathbf{r},t,\mathbf{\alpha }),
\label{hidden-variable fluid fields}
\end{equation}%
$\mathbf{\alpha }\in V_{\mathbf{\alpha }}\subseteq
%TCIMACRO{\U{211d} }%
%BeginExpansion
\mathbb{R}
%EndExpansion
^{k}$ (with $k\geq 1$) denoting a suitable set of \textit{hidden variables.}
In fluid dynamics these are intended as:

\subsubsection{\textbf{Definition - Hidden variables}}

A \textit{hidden variable} is as an arbitrary real variable which is
independent of\textit{\ }$(\mathbf{r},t)$ and is not an observable$.$

\subsection{Deterministic and stochastic fluid fields}

Hence, fluid fields of the type (\ref{hidden-variable fluid fields}) are
manifestly non-observables. However, if in the whole set $\overline{\Omega }%
\times I\times V_{\mathbf{\alpha }},$ they are uniquely-prescribed functions
of $(\mathbf{r},t,\mathbf{\alpha })$ then they are \textit{conditional
observables}. Hidden variables can be considered in principle either \textit{%
deterministic} or as \textit{stochastic variables, }in the sense specified
as follows.

\subsubsection{\textbf{Definition - Stochastic variables}}

Let $(S,\Sigma ,P)$ be a probability space; a measurable function $\mathbf{%
\alpha :}S\longrightarrow V_{\mathbf{\alpha }}$, where $V_{\mathbf{\alpha }%
}\subseteq
%TCIMACRO{\U{211d} }%
%BeginExpansion
\mathbb{R}
%EndExpansion
^{k}$, is called \textit{stochastic} (or \textit{random}) \textit{variable}.

A stochastic variable $\mathbf{\alpha }$\ is called \textit{continuous} if%
\textit{\ }it is endowed with a \textit{stochastic model} $\left\{ g_{%
\mathbf{\alpha }},V_{\mathbf{\alpha }}\right\} ,$\textit{\ }namely a real
function\textit{\ }$g_{\mathbf{\alpha }}$ (called as \textit{stochastic PDF})%
\textit{\ }defined on the set $V_{\mathbf{\alpha }}$ and such that:

1) $g_{\mathbf{\alpha }}$ is measurable, non-negative, and of the form
\begin{equation}
g_{\mathbf{\alpha }}=g_{\mathbf{\alpha }}(\mathbf{r},t,\mathbf{\cdot });
\label{stochastic PDF}
\end{equation}

2) if $A\subseteq V_{\mathbf{\alpha }}$ is an arbitrary Borelian subset of $%
V_{\mathbf{\alpha }}$ (written $A\in \mathcal{B}(V_{\mathbf{\alpha }})$),
the integral
\begin{equation}
P_{\mathbf{\alpha }}(A)=\int\limits_{A}d\mathbf{x}g_{\mathbf{\alpha }}(%
\mathbf{r},t,\mathbf{x})  \label{dist-of-alpha}
\end{equation}%
exists and is the probability that $\mathbf{\alpha \in }A$; in particular,
since $\mathbf{\alpha }\in V_{\mathbf{\alpha }}$, $g_{\mathbf{\alpha }}$
admits the normalization
\begin{equation}
\int\limits_{V_{\mathbf{\alpha }}}d\mathbf{x}g_{\mathbf{\alpha }}(\mathbf{r}%
,t,\mathbf{x})=P_{\mathbf{\alpha }}(V_{\mathbf{\alpha }})=1.
\label{normalization}
\end{equation}

The set function $P_{\mathbf{\alpha }}:\mathcal{B}(V_{\mathbf{\alpha }%
})\rightarrow \lbrack 0,1]$ defined by (\ref{dist-of-alpha}) is a
probability measure and is called distribution (or law) of $\mathbf{\alpha }$%
. Consequently, if a function $f\mathbf{:}V_{\mathbf{\alpha }%
}\longrightarrow V_{f}\subseteq
%TCIMACRO{\U{211d} }%
%BeginExpansion
\mathbb{R}
%EndExpansion
^{m}$ is measurable, $f$ is a stochastic variable too.

Finally define the \textit{stochastic-averaging operator }$\left\langle
\cdot \right\rangle _{\mathbf{\alpha }}$(see also \cite%
{Tessarotto2008-4,Tessarotto2009-1}) as\textit{\ }%
\begin{equation}
\left\langle f\right\rangle _{\mathbf{\alpha }}=\left\langle f(\mathbf{y}%
,\cdot )\right\rangle _{\mathbf{\alpha }}\equiv \int\limits_{V_{\mathbf{%
\alpha }}}d\mathbf{x}g_{\mathbf{\alpha }}(\mathbf{r},t,\mathbf{x})f(\mathbf{y%
},\mathbf{x}),  \label{stochastic averaging operator}
\end{equation}%
for any $P_{\mathbf{\alpha }}$-integrable function $f(\mathbf{y},\cdot ):V_{%
\mathbf{\alpha }}\rightarrow
%TCIMACRO{\U{211d} }%
%BeginExpansion
\mathbb{R}
%EndExpansion
$, where the vector $\mathbf{y}$ is some parameter.

\subsubsection{\textbf{Definition - Homogeneous, stationary stochastic model}%
}

The stochastic model $\left\{ g_{\mathbf{\alpha }},V_{\mathbf{\alpha }%
}\right\} $ is denoted:

a) \textit{homogeneous }if $g_{\mathbf{\alpha }}$ is independent of $\mathbf{%
r,}$ namely
\begin{equation}
g_{\mathbf{\alpha }}=g_{\mathbf{\alpha }}(t,\mathbf{\cdot });
\label{homogeneous stoch PDF}
\end{equation}

b) \textit{stationary} if $g_{\mathbf{\alpha }}$ is independent of $t,$
i.e.,
\begin{equation}
g_{\mathbf{\alpha }}=g_{\mathbf{\alpha }}(\mathbf{r},\mathbf{\cdot }).
\label{stationary stoch PDF}
\end{equation}

\subsubsection{\textbf{Definition - Deterministic variables}}

Instead, if $g_{\mathbf{\alpha }}(\mathbf{r},t,\mathbf{\cdot })$ is a
\textit{deterministic PDF}, namely it is of the form%
\begin{equation}
g_{\mathbf{\alpha }}(\mathbf{r},t,\mathbf{x})=\delta ^{(k)}(\mathbf{x-\alpha
}_{o}),  \label{deterministic g}
\end{equation}

$\delta ^{(k)}(\mathbf{x-\alpha }_{o})$ denoting the $k$-dimensional Dirac
delta in the space $V_{\mathbf{\alpha }},$ the hidden variables $\mathbf{%
\alpha }$ are denoted as \textit{deterministic}.

Let us now assume that, for a suitable stochastic model $\left\{ g_{\mathbf{%
\alpha }},V_{\mathbf{\alpha }}\right\} $, with $g_{\mathbf{\alpha }}$
non-deterministic, the stochastic variables $Z_{i}\equiv Z_{i}(\mathbf{r},t,%
\mathbf{\alpha })$ a{nd } $f_{1}(\mathbf{r,v},t,\mathbf{\alpha })$ (where $%
Z_{i}(\mathbf{r},t,\mathbf{\cdot })$ a{nd } $f_{1}(\mathbf{r,v},t,\mathbf{%
\cdot })${\ are measurable functions) a}dmit everywhere in $\overline{\Omega
}\times I$ and $\overline{\Gamma }\times I$ $\ $the \textit{stochastic
averages }$\left\langle Z_{i}\right\rangle _{\mathbf{\alpha }}$ and $%
\left\langle f_{1}\right\rangle _{\mathbf{\alpha }}$ defined by (\ref%
{stochastic averaging operator}).

Hence, $Z_{i}\equiv Z_{i}(\mathbf{r},t,\mathbf{\alpha }),$ $f_{1}(\mathbf{r,v%
},t,\mathbf{\alpha })$ and the mean-field force $\mathbf{F}(f_{1})$\textbf{\
}[see Sections 2,3 and 4]\textbf{\ }admit also the \textit{stochastic
decompositions}%
\begin{eqnarray}
Z_{i} &=&\left\langle Z_{i}\right\rangle _{\mathbf{\alpha }}+\delta Z_{i},
\label{stochastic decomposition} \\
f_{1} &=&\left\langle f_{1}\right\rangle _{\mathbf{\alpha }}+\delta f_{1},
\label{stochastic decomposition 2} \\
\mathbf{F}(f_{1}) &=&\left\langle \mathbf{F}(f_{1})\right\rangle _{\mathbf{%
\alpha }}+\delta \mathbf{F}(f_{1}).  \label{stochastic decomposition 3}
\end{eqnarray}

In particular, unless $g_{\mathbf{\alpha }}(\mathbf{r},t,\mathbf{\cdot })$
is suitably smooth, it follows that generally $\left\langle
Z_{i}\right\rangle _{\mathbf{\alpha }},\delta Z_{i}$ and respectively $%
\left\langle f_{1}\right\rangle _{\mathbf{\alpha }},\delta f_{1}$ may belong
to different functional classes with respect to the variables $(\mathbf{r}%
,t) $.

\subsection{Deterministic and stochastic INSE problems}

Therefore, assuming, for definiteness, that all the fluid fields $Z,$ the
volume force $\mathbf{f}$ and the initial and boundary conditions, are
either deterministic or stochastic variables and both belong to the same
functional class, i.e., are suitably smooth w.r. to $(\mathbf{r},t)$ and $%
\mathbf{\alpha }$,\ Eqs. (\ref{1b})- \ref{1c}) define respectively a \textit{%
deterministic} or \textit{stochastic initial-boundary value INSE problem}.
In both cases we shall assume that it admits a strong solution in $\overline{%
\Omega }\times I$ (or $\overline{\Omega }\times I\times V_{\alpha }$).

In the first case, which characterizes flows to be denoted as \textit{regular%
}, the fluid fields are by assumption \textit{physical observables,} i.e.,
uniquely-defined, smooth, real functions of $\ (\mathbf{r,}t)\in $ $\Omega
\times I$ \ [with $\Omega ,$ the \textit{configuration space,} and\textit{\ }%
$\overline{\Omega }$ its closure, to be assumed subsets of the Euclidean
space on\textit{\ }$%
%TCIMACRO{\U{211d} }%
%BeginExpansion
\mathbb{R}
%EndExpansion
^{3}$ and $I,$ the \textit{time axis,} denoting a subset of $%
%TCIMACRO{\U{211d} }%
%BeginExpansion
\mathbb{R}
%EndExpansion
$].

In the second case, characterizing instead \textit{turbulent
flows}, the fluid fields are only \textit{conditional observables}
(see again Subsection A). In this case, besides $\left(
\mathbf{r},t\right) $, they may be assumed to depend also on a
suitable stochastic variable $\mathbf{\alpha }$, (with
$\mathbf{\alpha \in }V_{\mathbf{\alpha }}$ and $V_{\mathbf{\alpha
}}$
subset of $%
%TCIMACRO{\U{211d} }%
%BeginExpansion
\mathbb{R}
%EndExpansion
^{k}$ with $k\geq 1$).\ Hence they are stochastic variables too.

\end{document}